\documentclass[useAMS,usenatbib,letter]{mn2e}
\usepackage{psfig} 

\def\xmm{{\sl XMM-Newton}}
\def\mcg{{MCG--6-30-15}}
\def\ngc{{NGC~4051}}

\title[Fluctuating-Accretion Model for AGN and BHXRB variability]{Investigating a Fluctuating-accretion Model for the Spectral-timing Properties of Accreting Black Hole Systems}  
\author[P. Ar\'evalo and P. Uttley]{P. Ar\'evalo$^{1}$\thanks{E-mail: parevalo@mpe.mpg.de} and P. Uttley$^{2}$\\ 
$^1$Max-Planck-Institut f\"ur Extraterrestrische Physik, Postfach 1312, D-85741 Garching, Germany\\ 
$^2$NASA Goddard Space Flight Center, Greenbelt, MD 20771, USA}

\begin{document}
\date{Received /Accepted}
\pagerange{\pageref{firstpage}--\pageref{lastpage}} \pubyear{2005}

\maketitle
\label{firstpage}

\begin{abstract} 
The fluctuating accretion model of \citet{Lyub} and its extension by
\citet{Kotov}, seeks to explain the spectral-timing properties of the
X-ray variability of accreting black holes in terms of
inward-propagating mass accretion fluctuations produced at a broad
range of radii. The fluctuations modulate the X-ray emitting region as
they move inwards and can produce temporal-frequency-dependent lags
between energy bands, and energy-dependent power spectral densities
(PSDs) as a result of the different emissivity profiles, which may be
expected at different X-ray energies.  Here we use a simple numerical
implementation to investigate in detail the X-ray spectral-timing
properties of the model and their relation to several physically
interesting parameters, namely the emissivity profile in different
energy bands, the geometrical thickness and viscosity parameter of the
accretion flow, the strength of damping on the fluctuations and the
temporal coherence (measured by the `quality-factor', $Q$) of the
fluctuations introduced at each radius.  We find that a geometrically
thick flow with large viscosity parameter is favoured, and confirm
that the predicted lags are quite robust to changes in the emissivity
profile, and physical parameters of the accretion flow, which may help
to explain the similarity of the lag spectra in the low/hard and
high/soft states of Cyg~X-1.  We also demonstrate the model regime
where the light curves in different energy bands are highly spectrally
coherent.  We compare model predictions directly to X-ray data from
the Narrow Line Seyfert~1 galaxy \ngc\ and the BHXRB Cyg X-1 in its
high/soft state and show that this general scheme can reproduce
simultaneously the time lags and energy-dependence of the PSD.

\end{abstract}


\section{Introduction}

A common characteristic in Active Galactic Nuclei (AGN) and X-ray
binary systems (XRBs) is their strongly variable X-ray emission. Most
of the variability power is in the form of aperiodic fluctuations that
span several orders of magnitude in temporal frequency
\citep[e.g.][]{vanderKlis,McHardy4051}.  Rapid, large amplitude X-ray
variations, on time-scales down to milliseconds in XRBs and minutes in
AGN, are commonly observed
\citep[e.g.][]{RevHifreq,GierlinFlares,McHardy4051}, supporting the
expectation from accretion theory that the bulk of the emission must
originate close to the central black hole.  However, in many sources
(in both AGN and XRBs), large-amplitude X-ray variations, of tens of
per cent fractional rms, are observed over several decades of
time-scales \citep[e.g.][]{ReigCygX1,McHardy4051,Uttley3227},
including values orders of magnitude longer than the viscous
time-scale of the inner disc, suggesting that these fluctuations must
originate at large radii. This discrepancy motivated the appearance of
propagating-fluctuation models, as proposed by \citet{Lyub}, who noted
that while the \emph{emission} might be produced only in the innermost
regions, the \emph{variability} can originate throughout the accretion
flow. In this model, accretion rate fluctuations arise over a wide
range of radii and, correspondingly, a wide range of time-scales, and
propagate inwards to modulate the central X-ray emission.

As noted by \citet{Uttley} the model of \citet{Lyub} is consistent with the
observed linear relation between rms-variability amplitude and
mean count rate in XRB and AGN X-ray light curves, which implies a
coupling between fluctuations on different time-scales, ruling out
models where the variability arises from strictly independent flares
or active regions \citep{UttleyNonlin}.  Fluctuating accretion
models, on the other hand, are consistent with this relation: the
fluctuations couple together as they propagate down to the centre, so
low frequency fluctuations produced at large radii modulate higher
frequency fluctuations produced further in.  A linear rms-flux
relation is also observed in neutron star XRB systems, where the
emission mechanism is most likely different to the one operating in
black-hole systems \citep{Uttley_sax,Uttley}. 
This result provides additional evidence for
fluctuating-accretion scenarios because it suggests that the
variability originates in the accretion flow and not just locally via
the mechanism producing the emission \citep{Uttley_sax}.

Besides explaining the broad range of variability time-scales and the
rms-flux relation, the fluctuating accretion model can also explain
the spectral-timing properties of the variability, as noted by
\citet{Kotov}.  For example, it has long been known
\citep[e.g.][]{Miyamoto,NowakGX339} that X-ray variations in black
hole X-ray binaries (BHXRBs) often show hard lags, i.e. a delay in
hard X-ray variations with respect to soft X-rays, which is larger for
variations on longer time-scales, and at higher energies.  
The magnitudes of the
lags are typically of order one per cent of the variability
time-scale.  Similar
time-scale-dependent, hard lags have recently been discovered in AGN,
albeit on much longer time-scales, commensurate with their higher
black hole masses \citep[e.g.][]{Papadakis,VaughanMCG,McHardy4051}.
In their analytical extension to Lyubarskii's model, \citet{Kotov}
explain these lags by invoking a radially extended X-ray emitting
region with an energy-dependent profile.  In this scenario, the
response of the emission to a fluctuation in the accretion flow is a
function of the inward propagation time-scale, and hence radius of
origin, of the fluctuation, combined with the emissivity profile.
Hence, if the emissivity profile is more centrally concentrated at
higher energies, hard-band lags are produced such that the lags are
larger for longer time-scale variations.  \citet{Kotov} show that the
same basic picture can also explain the energy dependence of the PSD
of BHXRBs and AGN, where there is relatively more high-frequency power
observed at higher energies than at lower energies
\citep[e.g.][]{Nowak_lags,McHardy4051}. This is because the emitting
region acts as a low-pass filter and, as the emitting region of the
soft band is more extended, the higher frequency variations are
filtered more strongly in the soft band.

Due to its success in explaining many aspects of the X-ray variability
data, the fluctuating accretion model of \citet{Lyub} and
\citet{Kotov} warrants a deeper investigation, which is the aim of
this paper.  As shown by the brief analytical treatment of
\citet{Kotov}, in principle, the model produces time-scale-dependent
lags of approximately the same amplitude as observed in the data, and
energy-dependent PSDs.  However it is not clear exactly how the lag
spectrum (lag as a function of Fourier frequency) and energy
dependence of the PSD varies as a function of emissivity profile, or
the parameters of the accretion flow (assuming the standard disc model
of \citealt{Shakura} e.g. the viscosity parameter $\alpha$ or the
scale-height of the flow), or due to the effects of radial damping of
variations in the accretion flow.  Furthermore, since \citet{Kotov}
analytically determined spectral-timing properties by making the
simplifying assumption that the perturbation introduced into the
accretion flow at each radius is a  delta-function in time and
radius, it is important to determine the effects on spectral-timing
properties of a more realistic model, where  stochastic variations
over a broader range of time-scales are introduced at each radius.
Finally, the fluctuating accretion model should, in principle,
produce light curves which are highly correlated in different bands 
(i.e. spectrally coherent, see \cite{Vaughan_coh} and Appendix~B of
this
 paper), but this aspect of the model has not yet been studied.
To study these various effects in more detail, in
Section~\ref{model} we introduce a computational toy model for a
fluctuating accretion flow, based on the work of \citet{Lyub} and
\citet{Kotov}, and explore the dependence of the PSD, time lags and
coherence on the model parameters in  Section~\ref{properties}. We
show how the model predictions fit X-ray data from AGN and the BHXRB
Cyg~X-1 in Sections~\ref{agn} and \ref{xrb}.  Specifically, we will
concentrate on explaining the spectral-timing properties of the BHXRB
Cyg~X-1 in its high/soft state, and AGN which show similar variability
properties, since these states show rather simple $1/f$-type PSD
shapes, without complex quasi-periodic oscillations (QPOs). We discuss
the implications of our results in Section~\ref{discussion} and
summarise our results in Section~\ref{conclusions}.

\section{The Model}
\label{model}
\subsection{Model construction and basic assumptions}

The model is based on the scenario proposed by \citet{Lyub}, where
fluctuations propagate inward through the accretion flow, modulating
the emission of the inner regions. The fluctuations are produced on
time-scales related to the viscous time-scale at the radius of origin
and are uncorrelated for different radial scales.  \citet{Churazov} noted
that, in this class of models, the accretion rate fluctuations can be
carried efficiently, and be produced up to high Fourier frequencies,
by a geometrically thick accretion flow, identifying it with an
extended corona over the thin disc. 

We use standard accretion disc \citep{Shakura} considerations to
relate the fluctuation time-scales and propagation speeds to the
radial position in the disc, since the same relations are applicable
to geometrically thick and optically thin accretion flows, as well as
the standard geometrically thin disc.
Following \citet{Lyub}, we assume that each
independent annulus produces a pattern of fluctuations $\dot m(r,t)$,
having most of the variability power at the local viscous frequency
$f_{\rm visc}(r)=r^{-3/2}(H/R)^2 \alpha/2\pi$ \citep[e.g][]{Kato},
where $(H/R)$ is the disc scale height to radius ratio, $\alpha$ is
the viscosity parameter, the radial position $r$ is in units of
gravitational radii $R_{\rm g}=GM/c^2$ and the frequency is given
in terms of $c/R_{\rm g}$.
 The local accretion rate at any radius in the disc is allowed to
fluctuate around a value $\dot M_o(r,t)$, as $\dot M(r,t)=\dot
M_o(r,t)\times(1+\dot m (r,t))$, where $\dot m(r,t) <<1$. As this new
value of the accretion rate propagates inward, it serves as the $\dot
M_o(r,t)$ for the fluctuations produced further in, modulating the
higher frequency fluctuations. The propagation speed is assumed to
equal the local radial drift velocity $v_{\rm visc}(r)= r^{-1/2}
(H/R)^2\alpha$.

A key feature of this model is that the absolute amplitude of
accretion rate variations is proportional to the local accretion rate,
$\dot M_o(r,t)$, at every radius. Considering that the characteristic
frequencies of the input fluctuation $\dot m (r,t)$ decrease with
radius, this implies that the amplitude of high frequency fluctuations
is proportional to the actual value of the accretion rate, given by
fluctuations on longer time-scales, and this produces the rms-flux relation.

To generate light curves, following \citet{Kotov}, we assume that the
X-rays are emitted by a radially (but not necessarily vertically)
extended region. This emitting region might correspond to an accreting
corona, possibly sandwiching an optically thick, geometrically thin
disc, or be another type of optically thin accretion flow (e.g. ADAF,
\citealt{NarayanYi}).  Having no pre-defined emission mechanism we
simply assume that the emitted flux per unit area is proportional to
the local accretion rate, which introduces the variability, and to a
stable radial emissivity profile. The emissivity profile, $ \epsilon
(r)$, is taken to follow the radial rate of gravitational energy loss
in the accretion disc, $\epsilon (r) = r^{-3}(1-\sqrt{r_{\rm
min}/r})$, where $r_{\rm min}$ is the inner radius of the disc (and of
the emission), which we will fix at $6 R_{\rm g}$ in our
simulations. This profile describes the \emph{total} energy
loss. However, as the emitted spectrum might be radially dependent, a
given energy band can have a different emissivity profile and so we
will use $\epsilon (r) = r^{-\gamma}(1-\sqrt{r_{\rm min}/r})$, where
the emissivity index $\gamma$ is a model parameter.

\subsection{Numerical implementation} 

For the calculation of the variability patterns, we discretize
the models of \citet{Lyub} and \citet{Kotov} by assuming that a
finite number of annuli produce independent fluctuations. These annuli
are spaced keeping a constant ratio between consecutive radii, to
mimic the behaviour of a continuous accretion flow that produces equal
variability power at each radial scale. This geometric spacing of the
radii introduces equal power per decade in time-scale, reproducing the
$1/f$ PSD shape, as predicted by \citet{Lyub}.

All these independent annuli are assumed to produce a pattern of small
accretion rate fluctuations, $\dot m(r,t)$. To meet the
requirement of producing most of the variability power at the viscous
frequency, these signals are modelled as random fluctuations with a
Lorentzian-shaped PSD of variable width, with peak frequency $f_{\rm
peak}$ \footnote{The Lorentzian peak frequency, corresponds to the
frequency where the contribution to total rms is at its maximum, and
is related to the resonance or centroid frequency $f_{\rm centroid}$
as $f_{\rm peak}=f_{\rm centroid}\sqrt{1+1/4Q^2}$ (e.g. see
\citealt{Pottschmidt03}).  set to equal $f_{\rm visc}$}.  The width
of the Lorentzian is governed by the quality factor $Q$, equal to the
ratio of Lorentzian peak frequency to the full width at half
maximum. We used broad input PSDs, with $Q=0.5$, a value similar
to the four-Lorentzian decomposition of the PSD of Cyg~X-1 in the
low/hard state \citep{Pottschmidt03} and also much narrower PSDs with
$Q=10$, to study the influence of this parameters on the final PSDs
and Cross Spectra.  The normalisation of these signal PSDs was chosen
to produce a final fractional variability of the output light curves
$F_{\rm var} \sim 10\% - 50\%$, by assigning each of the signals an
rms $\sigma_{\rm sig}= \sqrt{F_{\rm var}^{2}/N}$, where $N\sim 1000$
is the number of independent annuli. We used the method described by
\citet{Timmer} to generate the signal time series $\dot m(r,t)$,
i.e. by generating periodograms with the underlying signal PSD and the
correct statistical properties of a noise process and obtaining $\dot
m(r,t)$ through their inverse Fourier transform. The resulting $\dot
m(r,t)$ have zero mean and rms amplitude $\sim\sigma_{\rm sig}\ll 1$.

The accretion rate at a given annulus, $\dot M(r_i,t)$, is then
calculated iteratively as the product of the accretion rate at that
position, including its contribution to the fluctuation, $1+\dot
m(r_i,t)$, by the accretion rate at the annulus lying directly
outside, which includes the fluctuations from all outer annuli,

\begin{equation}
\dot M(r_i,t) = \dot M_o \prod^{i}_{j=0}1+\dot m(r_j,t),
\end{equation}

so that each $\dot M (r_i,t)$ is a flicker noise time series with
power law PSD of slope $\sim-1$ up to $f_{\rm visc}(r_i)$, and shows
a linear rms-flux relation.

The effect of the extended emitting region is accounted for when
calculating the \emph{light curves}. First, an emissivity index
$\gamma$ is assumed for each energy band. Then, the light curves are
constructed by adding the variability patterns $\dot M(r_i,t)$ from
all annuli, which are weighted by the flux originating from each
annulus (i.e. $2\pi r \delta r \epsilon (r)$, where $\delta r$ is the
width of the annulus), and time-shifted by the propagation time of
fluctuations from the outermost annulus to the present annulus.

The disc parameters are either kept constant or allowed to vary
radially as $(H/R)^2 \alpha (r)=C r^{-\beta}$, where C is a scaling
constant and $\beta$ is an index which governs the radial variation
(e.g. constant $H/R$ and $\alpha$ correspond to $\beta=0$). We will
normally use $C=0.3 \times 6.0^\beta$, to produce thick disc
parameters $(H/R)^2=1$ and $\alpha =0.3$ at the innermost
radius. Radially-varying disc parameters change the radial separation
of independent annuli but preserve the separation of independent
characteristic frequencies. Therefore, we fix the geometrically spaced
frequencies and use $(H/R)^2\alpha (r)$ to calculate the corresponding
radii and propagation speeds.

\begin{figure}
\psfig{figure=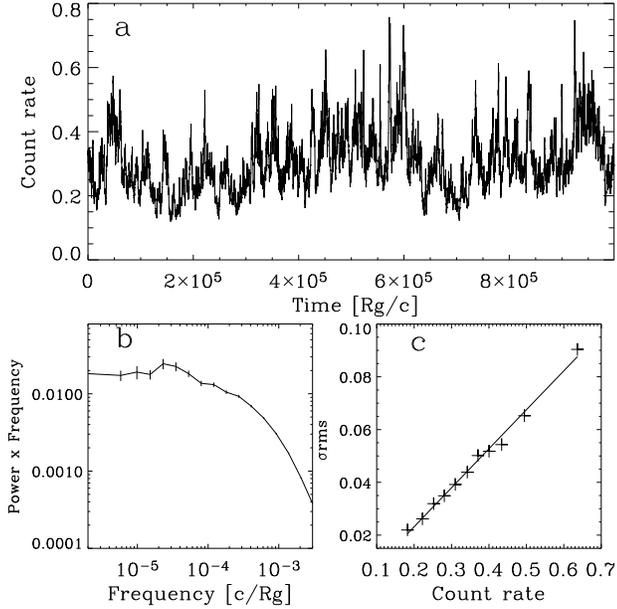,width=8.5cm,height=8.5cm} \caption {a:
Light curve from a realisation of the variability model, with
emissivity index $\gamma =3$. b: PSD of the same light curve, here
plotted as power $\times$ frequency, showing the typical 1/f shape
at low frequencies [i.e. a flat top in this representation, which
shows the relative contribution of rms from each decadal range in
frequency, analogous to the $\nu F(\nu)$ form of a spectral energy
distribution (SED)].  The PSD bends to steeper slopes towards
$f_{\rm visc}(r_{\min})=3\times 10^{-3} c/R_{\rm g}$. c:
$\sigma_{\rm rms}$ vs flux for the same light curve, showing a
linear relation.} 
\label{fig0}  
\end{figure}

Fig. \ref{fig0} shows a light curve realisation with emissivity index
$\gamma=3$ (a), together with its PSD, plotted as frequency $\times$
power (b), and its rms variability amplitude plotted vs the average
count rate of short sections of the light curve (c).

Finally, damping of the fluctuations as they propagate through the
accretion flow affects the power spectra of the transmitted
signals. The magnitude of the damping effect depends on the nature
of the fluctuations and their propagation mechanism, which are not
specified in our phenomenological model. In their analytical
model, \citet{Kotov} include this damping effect by using a Green
function for diffusion in the geometrically thin disc approximation.
Here, for simplicity and generality, we use a simple prescription
for damping by assuming that fluctuations of frequency $f$ are
significantly damped after travelling a distance $\Delta
r/r=\sqrt{f_{\rm visc}(r)/f}$, so that fluctuations on frequencies
much higher than the local viscous frequency are strongly suppressed
\citep[see e.g.][ but notice that in that case damping was 
specifically derived for diffusive propagation of a delta-function
shaped impulse in mass accretion rate]{Churazov}.  The damping
prescription is implemented by first taking the Fourier transform of
$\dot M(r_{i},t)$ at each radial step $r_{i}$, then multiplying the
real and imaginary coefficients of the Fourier transform at each
frequency $f$ by a frequency-dependent factor $exp \left( -D
\frac{\Delta r}{r_i}\sqrt{\frac{f}{f_{\rm visc}(r_i)}} \right)$
(where the damping coefficient D is a free parameter) and inverse
Fourier-transforming the result to get the new, damped $\dot
M(r_{i},t)$.

Throughout this paper we will use $R_{\rm g}=GM/c^2$
as units of distance and $R_{\rm g}/c$ as units of time. For a quick
check of frequencies and time-scales note that $R_{\rm g}/c \sim
(5M/M_6) s $ where $M_6=10^6 M_{\odot}$, so the frequencies quoted
would be in units of Hz for a $2 \times 10^5 M_{\odot}$ black hole
and units of $10^{4}$ Hz for a $20 M_{\odot}$ black hole. 

\subsection{Analytical estimates} 

The extended emitting region
acts as a low-pass filter on the PSD. This effect can be understood
by considering that the signals at low temporal frequencies are
imprinted in the flux from most of the emitting region, while higher
frequencies are only imprinted in the emission from smaller, inner,
areas. As the light curve has additive components from all these
regions, this produces comparatively less variability power at high
frequencies, creating a bend in the PSD. The `filtered' PSD can be
approximated by multiplying the original PSD($f$) by the squared
fraction of the total flux that is produced within the
characteristic radius $r_f$ of each frequency $f$ as

\begin{equation}  {\rm PSD}_{\rm filt}(f)={\rm PSD}(f)
\left(\frac{\int^{r_f}_{r_{\rm min}}\epsilon (r)2 \pi r
dr}{\int^{\infty}_{r_{\rm min}}\epsilon(r)2 \pi r dr}\right)^2
\end{equation}  

The filter factor tends to 1 as $r_f$ tends to
$\infty$, i.e. for low temporal frequencies the original variability
power is preserved, and decreases monotonically as $r_f$ reaches
$r_{\rm min}$, reducing the variability power at the high frequency
end. Obviously, as the emissivity profiles $\epsilon (r)$ used are
steep, it makes little difference to the filtered PSD if the
emission region is truncated at a large radius rather than extending
to infinity. 

The filtering effect due to the origin of the
signals within the emitting region could be easily calculated once
the emissivity profile is defined, and its effect is simply to
reduce the overall {\it normalisation} of the contribution to the
PSD from each signal (in a frequency-dependent way), without
distorting the signal PSD shape. However, the finite travel time of
the fluctuations through the emitting region distorts the PSD shape
of the input signals, by smoothing out high frequency fluctuations,
thus reducing the variability amplitude further, while damping can
also affect the low-frequency power. Therefore, the total effective
filtering of the underlying fluctuations is quite complex, so we
will primarily use numerical simulations to demonstrate its effects
in the following section. 

The frequency-resolved time lags can be approximated by defining, for
each signal $s$ (produced at radius $r_s$) an `average travel
time', $\bar{\tau }(r_s)$, by weighting the travel time of the
fluctuations  from $r_s$ to the radius $r$, $\tau(r,r_s)$ by the
emissivity profile $\epsilon (r)$,

\begin{eqnarray} 
\label{eqlags}
\bar{\tau}(r_s)=\frac{\int^{r_s}_{r_{\rm min}}\tau(r,r_s) \epsilon
(r)
 2\pi r dr}{\int^{r_s}_{r_{\rm min}} \epsilon(r) 2 \pi r dr},\\
 {\rm where}\qquad \tau (r,r_s)=\int_{r}^{r_s}{\frac{d\tilde r}{v_{\rm visc}(\tilde r)}}.\nonumber
\end{eqnarray}

If the signals are simple fluctuations on the viscous time-scale only
(i.e. defined by a delta-function and not a broad Lorentzian PSD),
then $r_s$ can be expressed in terms of frequency and this formula
gives the mean travel time as a function of $f$. Otherwise, if the
signals are characterised by broad Lorentzian PSDs, a given temporal
frequency will have important contributions from a range in radii
$r_s$. In this case, the mean travel time at a given frequency $f$ is
a sum of $\bar{\tau}$ for each contributing (i.e. overlapping)
Lorentzian signal PSD, weighted by the variability power at $f$
contributed by that signal.

Time lags between two energy bands appear when they are characterised
by different emissivity indices, since this produces different mean
travel times for each band. We will use the difference of $\bar \tau
(f)$ calculated with the corresponding $\gamma$ indices as an
approximation to the time lags between energy bands.

An explicit form for the filter factor of the PSD and for the time lags is
given in Appendix~A.

\section{Spectral-Timing Properties} 
\label{properties}

The model was constructed to produce power law-PSD time series that
also reproduce the non-linear properties of AGN and BHXRB light
curves. The exact PSD shape and spectral-timing properties between
different energy bands, depend on a few key model parameters:
emissivity indices, disc structure parameters, PSD of the input
signals and damping coefficient. Here we will explore how each of
these parameters affect the observable properties.  An overview
of the spectral-timing measurements we use here is provided in
Appendix~B.

\subsection{Dependence on emissivity indices}

The loss of gravitational energy in the accretion disc requires an
energy release per unit area proportional to $\epsilon (r)=
r^{-3}(1-\sqrt{r_{\rm min}/r})$ \citep{Shakura}. A given energy band
will have the same radial emissivity profile only if it represents a
constant fraction of the emitted flux at each radius.  However,
following \citet{Kotov}, we assume the emitted spectrum hardens
towards the centre, so the actual radial emissivity profile will be
steeper for harder bands. We will refer to time series constructed
with emissivity index $\gamma=3$ as the `soft' light curve and use
higher values of $\gamma$ for the `hard' bands.  In this subsection,
we assume the product of disc structure parameters
$(H/R)^{2}\alpha=0.3$, and that there is no damping ($D=0$).

\begin{figure} 
\psfig{figure=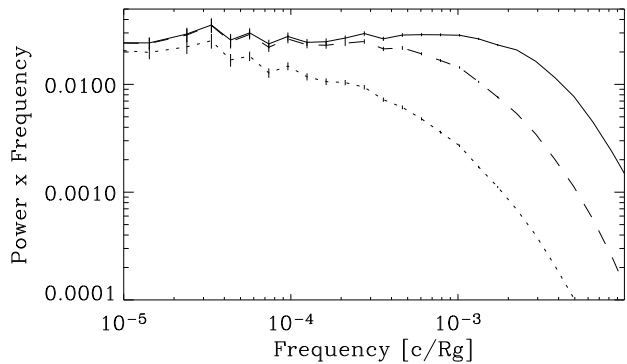,width=8.5cm,height=5cm}
\caption{Filtering effect of an extended emitting region. The PSD of the
variability pattern at the innermost radius, shown in the solid line,
keeps a $1/f$ shape up to $f_{\rm visc}(r_{\rm min})$. `Soft'
($\gamma=3$) and `hard' ($\gamma=5$) light curves of the same
realisation, in dotted and dashed lines respectively, lose variability
power at high Fourier frequencies due to the effect of the
extended emitting region. For this realisation we used model parameters $Q=0.5,\beta=0,D=0$.} 
\label{1a_per} 
\end{figure}

The PSD of the final variability pattern at $r_{\rm min}$ for a
realisation with input signal $Q=0.5$, is shown by the solid line in
Fig.\ref{1a_per}. This PSD would correspond to an infinitely
concentrated emitting region located at $r_{\rm min}$, or
equivalently, a compact emitting region located entirely {\it within}
$r_{\rm min}$. It keeps a $1/f$ behaviour (corresponding to a flat top
in this plot of power multiplied by frequency) up to $f_{\rm
visc}(r_{\rm min})=3 \times 10^{-3} c/R_{\rm g}$, as no annuli
contribute with significant variability power above this frequency,
the PSD bends downward. The dotted and dashed lines show the PSDs of
soft and hard light curves from a realisation with the same underlying
accretion fluctuations, with $\gamma_{\rm soft} =3$ and $\gamma_{\rm
hard} = 5$, respectively. The figure highlights the filtering effect
of the extended emission on these two light curves, which makes them
lose variability power at high frequencies, shifting the bend to lower
frequencies. On longer time-scales, all the PSDs flatten to their
original slope of -1. The resulting PSDs bend continuously to steeper
slopes at high frequencies but, for a limited frequency range, they
can be approximated fairly well by a singly bending power law model, as
has been used to fit real data. Following e.g. \citet{McHardy4051}, we
will use the following bending power law model to fit the PSD,

\begin{equation}
\label{eq1}
P(f)=Af^{\alpha_L}\left(1+\left(\frac{f}
{f_{\rm b}}\right)^{\alpha_L-\alpha_{\rm
H}}\right)^{-1} .
\end{equation}

Table \ref{t1} gives the values for the low and high frequency slopes,
$\alpha_{\rm L} $ and $\alpha_{\rm H}$ respectively, and the break
frequencies $f_{\rm b}$ that result from fitting Equation~\ref{eq1} to
PSDs with emissivity indices $\gamma$= 3, 4 and 5, for broad ($Q=0.5$)
and narrow ($Q=10$) input signal PSDs. These values depend slightly on
the frequency range used for fitting, but reveal the general trend of
more high-frequency power for harder energy bands, given by flatter
$\alpha_{\rm L}$, and higher $f_{\rm b}$ and/or flatter $\alpha_{\rm
  H}$. Note that the underlying PSDs bend continuously to steeper
slopes at higher frequencies, so limiting the frequency range used for
fitting, excluding higher frequencies, produces flatter high-frequency
slopes.

\begin{table}
\label{t1}
\begin{tabular}{lcccc}
\hline
&$\gamma$&$\alpha_{\rm L}$& $f_{\rm b}$ ($\times 10^{-3} c/R_{\rm g}$)&$\alpha_{\rm H}$\\
\hline
$Q=0.5$&3&-1.48    & 1.27    &  -3.74\\
&4&-1.22     & 1.38    &  -3.54\\
&5&-1.10     & 1.54    &  -3.46\\
\hline
$Q=10$&3&-1.44   &   0.80   &   -4.73 \\
&4&-1.16   &   0.96   &   -4.79\\
&5&-1.02   &   1.12   &   -4.86\\

\hline 
\end{tabular} 
\caption{Bending power-law model fits to the PSDs
of simulated light curves, with emissivity indices $\gamma$= 3, 4 and 5,
for broad ($Q=0.5$) and narrow ($Q=10$) input signal PSDs. Higher $\gamma$
values, corresponding to higher energy bands, retain more high-frequency power, evident in the flatter low-frequency slope
$\alpha_{\rm L}$ and in and higher $f_{\rm b}$ and/or flatter
$\alpha_{\rm H}$. The fitting range used was $6\times
10^{-5}<f<10^{-2} c/R_{\rm g}$.  Note that the underlying PSDs bend continuously to steeper slopes at higher frequencies, so limiting the frequency range used for fitting, excluding higher frequencies, produces flatter slopes above the break.} 
\end{table}

\begin{figure} 
\psfig{figure=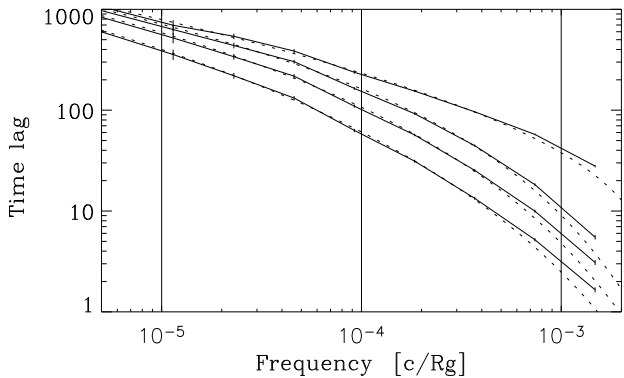,width=8.5cm,height=5.2cm}
\caption {Time lags as a function of Fourier frequency, between a soft light curve of emissivity index $\gamma =3$ and hard light curves of indices $\gamma = 3.5, 4, 5 $ and $\infty$, for $Q=10$. Larger lags correspond to bigger differences between emissivity indices. The  points joint by solid lines represent the values calculated from  numerical realisations of the light curves and the dotted lines represent the lags predicted by Eq. \ref{eqlags} using the corresponding model parameters. The vertical lines mark the frequencies that will be used in the following two figures to measure lags.}  
\label{lags_em} 
\end{figure}

The emissivity indices $\gamma$ also define the average travel time of
the signals as seen by the different energy bands, so they play a
major role in determining the time lags.  For emissivity profiles as
used here, the lags show an almost power law dependence on frequency,
as can be seen in Fig.~\ref{lags_em}. This figure shows, in solid
lines, the time lags between a soft light curve with $\gamma_{\rm
soft}= 3$ and hard light curves of indices $\gamma_{\rm hard}=$ 3.5,
4, 5 and $\infty$, calculated for $Q=10$. Obviously, this last light
curve, taken to be the variability pattern at $r_{\rm min}$, marks the
limiting lag spectrum that can be obtained by increasing $\gamma_{\rm
hard}$, assuming the other parameters are fixed at the values given
above.  The approximate slope of these lag spectra in the range
$10^{-5}-10^{-4} c/R_{\rm g}$ goes from to -0.8 for $\gamma_{\rm
hard}= 3.5$ to -0.5 for $\gamma_{\rm hard}= \infty$.

The dotted lines in Fig. \ref{lags_em} are the time lags predicted by
Eq. \ref{eqlags} using the parameters of the corresponding light curve
simulations. As these analytic estimates give a good fit to the
simulated data, we will use them hereafter on their own or to
complement the time lags calculated from simulations.

\begin{figure}
\psfig{figure=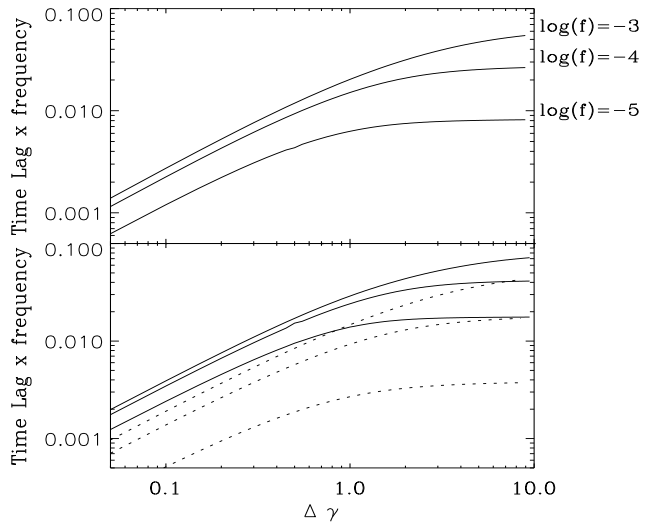,width=8.5cm,height=7cm}
\caption {Top panel: Fractional time lags (lags $\times$ frequency) for energy
bands of emissivity index \emph{difference} $\Delta \gamma= \gamma_{\rm
hard}-\gamma_{\rm soft}$, with $\gamma_{\rm soft}=3$, measured at three
temporal frequencies, $f=10^{-3}, 10^{-4}, 10^{-5} c/R_{\rm g}$. The lags
increase with the difference in emissivity indices and depend sensitively on
this parameter up to $\Delta \gamma \sim 1$, for larger differences the lags
remain at around $1\%-10\%$ of the Fourier frequency. As the slope of the lag
spectra is greater than -1, fractional lags are
smaller at low frequencies. Bottom panel: Same as
above but using $\gamma_{\rm soft}=2.5$ (solid lines) and $\gamma_{\rm
soft}=3.5$ (dashed lines), 
a steeper emissivity profile for the soft band produces smaller
lags for the same difference in indices. In both panels the lags were calculated for narrow input signal PSDs, with $Q=10$. The value of $Q$ affects mainly the lags at high frequencies, the lags at $10^{-5}$ and $10^{-4}c/R_{\rm g} $ in these plots are almost independent of the $Q$-value used.}
\label{lags_em2}
\end{figure}

The magnitude of the time lags increases with the difference between
the emissivity indices of the light curves, $\Delta \gamma=
\gamma_{\rm hard}-\gamma_{\rm soft}$. Fig.\ref{lags_em2} shows the
fractional lags (lag $\times$ frequency, i.e. equal to $\phi/2\pi$, where $\phi$ is the phase lag, see Appendix B) at
the three Fourier frequencies marked in Fig.\ref{lags_em} by the
vertical lines, $f=10^{-3}, 10^{-4}, 10^{-5} c/R_{\rm g}$, as a
function of $\Delta \gamma$, calculated for narrow input signal PSD,
$Q=10$. Each set of lines was calculated with a fixed value of
$\gamma_{\rm soft}$ and varying $\gamma_{\rm hard}$. In all cases, the
lags increase rapidly with $\Delta \gamma$ up to $\Delta \gamma \sim
1$. For larger differences in emissivity indices, the lags remain at
around $1\%-10\%$ of the variability time-scale, tending towards the
limiting value obtained for $\gamma_{\rm hard}=\infty$.  In this
figure, for the top panel we used $\gamma_{\rm soft}=3$ and for the
bottom panel, $\gamma_{\rm soft}=2.5$ (solid lines) and $\gamma_{\rm
soft}=3.5$ (dashed lines). As can be seen, for any $\Delta \gamma$,
the lags increase with decreasing $\gamma_{\rm soft}$, i.e. a more
extended soft emitting region naturally produces larger lags with
respect to all hard bands. Finally, the magnitude of the lags
increases with decreasing frequency, but the slope of the lag spectrum
is $>-1$. This translates into smaller fractional lags at lower temporal
frequencies.

\subsection{Disc structure parameters}

In our model, the product of
disc structure parameters $(H/R)^2\alpha$ can be treated as
a single term that governs simultaneously the characteristic
frequencies of each radius and the propagation speed. We do not attempt here
to model the evolution of the disc structure but only to explore the effect
that varying these parameters globally has on the observable quantities. We
use the prescription $(H/R)^2 \alpha=C r^{-\beta}$ to treat disc parameters
that vary radially. Assuming constant $\alpha$,
a positive value of $\beta$ implies that the accretion
flow flattens outwards, approximating the case of a thin disc that expands
into a thick disc or ADAF-type flow towards the centre. We have used
$(H/R)^2 \alpha=0.3$ at $r_{\rm min}$.  Assuming $\alpha=0.3$,
this corresponds to a geometrically thick disc.
The motivation for assuming this large value for $(H/R)^2 \alpha$
is the need to maximize the characteristic frequencies of the innermost
input signals, as the data can show significant variability up to the orbital
time-scale of the last stable orbit.

Varying $\beta$ changes the spacing of the radii that correspond to
independent characteristic frequencies, but not the spacing of the
frequencies themselves. The variability pattern at each independent
annulus is not affected, but the position of each annulus in the disc
shifts. Therefore, the final light curves are affected only through
the effects of the extended emitting region, as the emissivity of each
annulus and the travel time between them changes. The overall effect
on the PSDs is a small decrease in power towards high frequencies,
produced by a steeper low-frequency PSD slope (see Table
\ref{lfslope}).

\begin{figure}
\psfig{figure=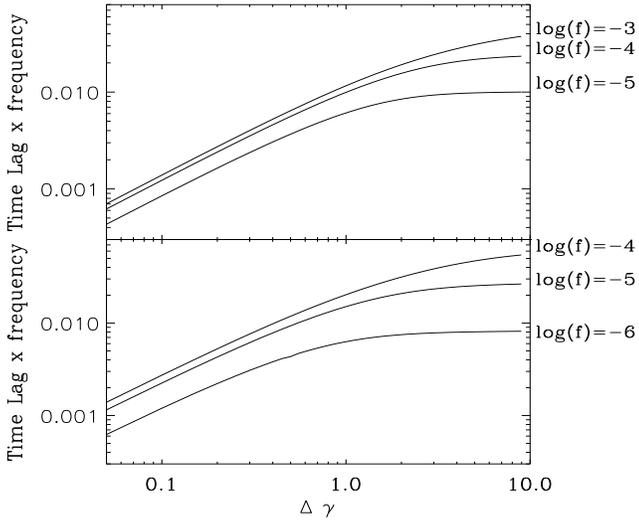,width=8.5cm,height=7cm}
\caption { Top panel: Same as top panel of Fig. \ref{lags_em} but using
$\beta=1$, i.e. constant $(H/R)^2\alpha$. The lags are similar to the
$\beta=0$ case but the slope of the lag spectrum is closer to -1, so
the fractional lags have approximately the same value at all Fourier
frequencies. Bottom panel: Same as top panel of Fig. \ref{lags_em} 
(using $\beta = 0$) but using $(H/R)^2 \alpha=0.03$.
 The lags are equivalent to the case with $(H/R)^2 \alpha=0.3$
(top panel in Fig. \ref{lags_em2}) but for frequencies a factor of 10 lower. In both panels the lags were calculated for narrow input signal PSDs, with $Q=10$.} 
\label{lags_em3}
\end{figure}

The amplitude of the time lags remains approximately unchanged,
for the frequency range studied ($10^{-3}-10^{-5} c/R_{\rm g}$), but the
slope of the lag spectra steepens. For $\beta=1$ and $\gamma_{\rm
soft}=3$, the lag spectrum slope in the frequency range
$10^{-5}-10^{-4} c/R_{\rm g}$ goes from -1.0 for $\gamma_{\rm
hard}=3.5$ to -0.7 for $\gamma_{\rm hard}=\infty$ (c.f. the equivalent
values for $\beta = 0$ of $-0.8$ and $-0.5$ respectively).  The top
panel in Fig. \ref{lags_em3}, shows the fractional lags as a function
of $\Delta \gamma$ for $\gamma_{\rm soft}=3$ and $\beta=1$. Evidently,
these functions are very similar to the $\beta=0$ case shown in the
top panel of Fig. \ref{lags_em2}, having only slightly lower lags at
high frequencies and higher lags at low frequencies.

Changing the scaling constant C to lower values, i.e. as in a thinner
disc, shifts all the time-scales up by the same factor. This effect
shifts the PSD to lower frequencies without changing its shape.  The
lower panel in Fig.\ref{lags_em3} shows the fractional time lags as a
function of $\Delta \gamma$ for $\beta=0$ and $(H/R)^2 \alpha= 0.03$
instead of $(H/R)^2 \alpha= 0.3$, used in all previous cases. The
fractional lag spectrum is equivalent to the case with $(H/R)^2
\alpha=0.3$ (top panel in Fig. \ref{lags_em2}) except shifted down a
factor of 10 in frequency, so the absolute value of the lag at a given
frequency is a factor of 10 higher than for $(H/R)^2 \alpha= 0.3$.
Notice however, that as the PSD also shifts in frequency, the
fractional lags at, for example, the break frequency remain
approximately the same.

\subsection{Input signals} 
\label{input}
So far we have mainly considered light curves whose input signals have
a Lorentzian PSD of quality factor $Q=10$. As this shape is quite
arbitrary, it is important to estimate the influence of the signal PSD on
the observable parameters.

\begin{figure}
\psfig{figure=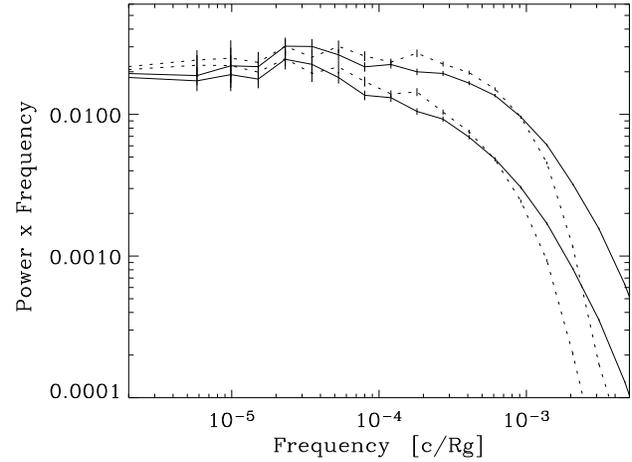,width=8.5cm,height=6.3cm}
\caption {PSDs of soft ($\gamma =3$, bottom) and hard ($\gamma =4$, top), of two model realisations, with input signal PSD quality factor $Q=0.5$ (in solid lines) and $Q=10$ (dotted lines). The light curves with $Q=10$ drop-off more sharply than their $Q=0.5$ counterparts close to the maximum frequency $f=3\times 10^{-3} c/R_{\rm g}$. At lower frequencies both realisations follow a $1/f$ relation.}
\label{input_psd}
\end{figure}

If the input signal PSDs are narrow, there is little variability power
above $ f_{\rm visc}(r_{\rm min})$ and the final PSD bends down
sharply above this frequency, whereas broader input signal PSDs spill power
up to higher frequencies making this drop less pronounced. Fig.
\ref{input_psd} shows the PSDs of soft and hard light curve pairs from
two realisations with different $Q$ values for the signal PSDs. 
For each emissivity index,
the broad- and narrow-input-PSD realisations, plotted in dashed and
solid lines respectively, have similar final power spectra, only
differing noticeably near $ f_{\rm visc}(r_{\rm min})=3\times 10^{-3}
c/R_{\rm g}$. The same results are obtained
from changing the signal PSD to, for
example, a doubly-broken power law. The $1/f$ behaviour of the low
frequency PSDs is produced by using equal rms per input signal
and geometric separation of the signal characteristic frequencies
(i.e. Lorentzian peak frequencies),
and does not depend on the shape of the signal PSDs. On the other hand,
the signal PSDs define the high frequency slope, so this part of the
final PSD can serve as a diagnostic for the intrinsic variability
process.

\begin{figure} 
\psfig{figure=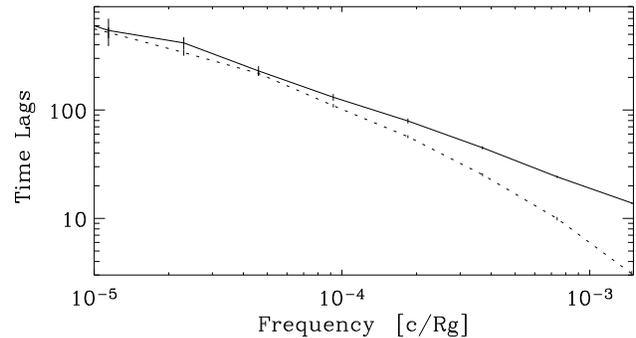,width=8.5cm,height=4.5cm}
\caption {Time lags between soft ($\gamma =3$) and hard ($\gamma =4$) light curve pairs with input-signal PSD quality factors $Q=0.5$ (solid line), $Q=10$ (dotted line). Broader-PSD signals
increase the time lags close to the maximum frequency.}  
\label{input_lags}
\end{figure}

The signal PSD also affects the high frequency end of the lag
spectra. Fig. \ref{input_lags} shows how time lags drop off less
sharply for a broad-PSD signal. At low frequencies,
however, there is no evident dependence of time lag on signal PSD.

\begin{figure}
\psfig{figure=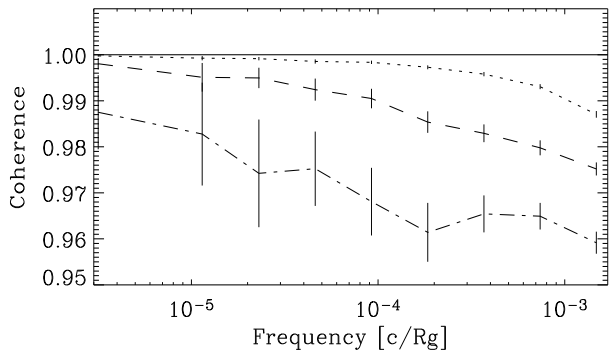,width=8.3cm,height=4.75cm}
\caption {Coherence between soft ($\gamma=3$) and hard ($\gamma=4$) light
 curve pairs, with Lorentzian input signal PSDs with quality-factors
$Q=10$ (dotted line), and $Q=0.5$ (dashed line). The light curves with broader
input-signal PSDs
show a stronger loss of coherence. The dot-dashed
 line corresponds to a $Q=0.5$ light curve pair affected by damping ($D=1$),
 which reduces the coherence by a few extra per~cent. For comparison,
 coherence = 1 is shown by the solid line.}
\label{input_coh}
\end{figure}

The signal PSD has a stronger effect on the coherence between two
energy bands. For broad signal PSDs, a given temporal frequency will
have contributions from several (incoherent) signals produced by
different annuli. As the different emissivity indices highlight
differing regions of the disc, the coherence between the corresponding
energy bands can drop. This is shown in Fig. \ref{input_coh}, where we
plot the coherence for light curves with broad ($Q=0.5$) and narrow
($Q=10$) signal PSDs, using dotted and dashed lines respectively. As
expected, broader signal PSDs produce a stronger drop. Notice however,
that in either case this loss of coherence is at most a few
per~cent. The emissivity indices have a much stronger effect on the
coherence. For Fig. \ref{input_coh} we used $\gamma_{\rm soft}=3$ and
$\gamma_{\rm hard}=4$. The same plot calculated for $\gamma_{\rm
soft}=3$ and $\gamma_{\rm hard}=5$ would show a drop of up to 7 per~cent, and
for $\gamma_{\rm soft}=3$ and $\gamma_{\rm hard}=\infty$, of 40 per~cent, for
light curves with $Q=0.5$.

\subsection{Damping}
\begin{figure}
\psfig{figure=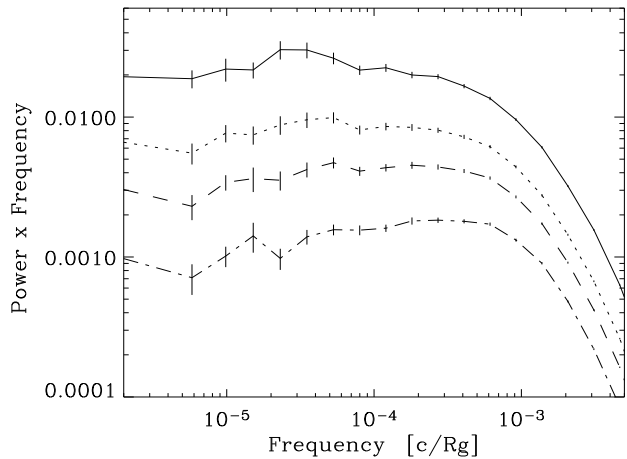,width=8.5cm,height=6.3cm}
\caption {Effect of damping on the PSD. The solid line shows the PSD of a
 ($\gamma =4$) light curve calculated without damping. The dotted, dashed and
 dot-dashed lines shows the PSDs of similar light curves calculated with
 damping coefficients D=0.5, 1 and 2, respectively. Damping reduces the
 overall power but acts more strongly on the lowest frequencies so the low
 frequency slope $\alpha_{\rm L}$ flattens.}
\label{damp_per}
\end{figure}

The effect of damping is to reduce the amplitude of the fluctuations as they
propagate inward so that the farther they go the smaller they get. An
immediate consequence is the reduction of variability power at all frequencies
as seen in the PSD. As this suppression is frequency dependent, damping can
change the shape of the PSD. Fig. \ref{damp_per} shows the PSD for four hard
light curves ($\gamma=4$) with equal disc parameters and input signal
variability
amplitudes, but affected by different degrees of damping. The most noticeable
change is the overall reduction of variability power. However, as the input
rms is arbitrary, this effect does not help to quantify the amount of damping
affecting real data. A more important effect is the change in the low
frequency slope, as damping can produce values of $\alpha_{\rm L}<-1$, as
seen in e.g. \mcg\ \citep{McHardyMCG}. Table \ref{lfslope} shows the values
of $\alpha_{\rm L}$ for damping coefficients $D =0,0.5,1$ and 2, for
different emissivity indices. In all cases, the stronger damping flattens the
low frequency PSD, by similar values
regardless of $\gamma$, e.g. for $D=2$ compared to $D=0$,
$\Delta \alpha_{\rm L} \sim 0.17 $ for $\beta=0$ and
$\Delta \alpha_{\rm L} \sim 0.1$ for $\beta=1$.
 
\begin{table}
\begin{tabular}{cccccc}
\hline
&&&\multicolumn{3}{c}{$\alpha_L$}\\
\hline
Model&D&$\beta$& $\gamma=3$&$\gamma=4$&$\gamma=5$\\
\hline
1&2&0 &-0.98&-0.83&-0.80\\
2&2&1&-1.19&-0.99&-0.90\\
3&1&0&-1.04&-0.89&-0.86\\
4&1&1&-1.24&-1.04&-0.95\\
5&0.5&0&-1.09&-0.94&-0.90\\
6&0.5&1&-1.26&-1.07&-0.98\\
7&0&0&-1.14&-1.00&-0.96\\
8&0&1&-1.29&-1.10&-1.02\\
\hline
\end{tabular}
\caption{\label{lfslope}Low Fourier frequency slopes, $\alpha_{\rm L}$, 
fit to PSDs of light curves generated with damping coefficients D and radial
dependence of the disc parameters given by $\beta$ as described in the
text, for emissivity indices $\gamma=3, 4$ and 5. The 
fits are carried out only for data below $f=-3 \times 10^{-4} c/R_{\rm g}$.}   
\end{table}

Damping can also reduce the coherence, especially in the case of broad
input signal PSDs. The dot-dashed line in Fig. \ref{input_coh}, shows
the coherence for light curves of emissivity indices $\gamma_{\rm
soft}=3$ and $\gamma_{\rm hard}=4$, $Q=0.5$ and damping coefficient
$D=1$. These light curves only differ from those corresponding to the
dashed line on the same plot by the effect of damping, which increases
the loss of coherence at the highest frequency bin from 2 to
4~per~cent. Again, a larger difference in emissivity indices makes
this drop more pronounced, and damping effects increase these losses
accordingly: for $\gamma_{\rm soft}=3$ and $\gamma_{\rm hard}=5$ the
coherence at the highest frequency bin drops by $\sim 12$~per~cent and
for $\gamma_{\rm soft}=3$ and $\gamma_{\rm hard}=\infty$, by
60~per~cent. Finally, damping can produce slightly steeper lag
spectra, reducing the lags at high frequencies. This effect is small:
damping coefficients of $D=0$ and $D=1$ produce lags differing by a
factor of $< 1.5$ at frequencies close to the measured break-frequency
$f_{b}$ and less at lower frequencies.
\section{Comparison with AGN X-ray light curves} 
\label{agn} 

AGN X-ray light curves from long monitoring campaigns typically show
power law PSDs with a slope $\sim-1$ at low frequencies, bending or
breaking at high frequencies to slopes $\sim -2$ or steeper
\citep[e.g.][]{Uttley2002, Markowitz, McHardy4051, McHardyMCG,
Uttley3227}. The high-frequency PSD is energy dependent, with higher
energy PSDs showing a higher break frequency and/or flatter slopes
above the break, depending on model assumptions
(e.g. \citealt{McHardy4051}).  Significant frequency-dependent time
lags have been measured in a few AGN [NGC~7469 \citep{Papadakis};
\mcg\ \citep{VaughanMCG}; NGC~4051 \citep{McHardy4051}; NGC~3783
\citep{Markowitz3783}]. In all cases, these lags increase with
time-scale and energy separation of the bands and show magnitudes of
up to a few percent of the variability time-scale. The measured
coherence is usually high ($>0.9$) up to about the break frequency,
sometimes dropping by tens of per cent above the break
\citep{VaughanMCG,McHardy4051}. This behaviour is qualitatively
consistent with the predictions of the fluctuating accretion model
outlined in Section~3.  Now we will compare the model predictions
directly to AGN data to check this consistency quantitatively.

\subsection{Time lags} 

\begin{figure}
\psfig{figure=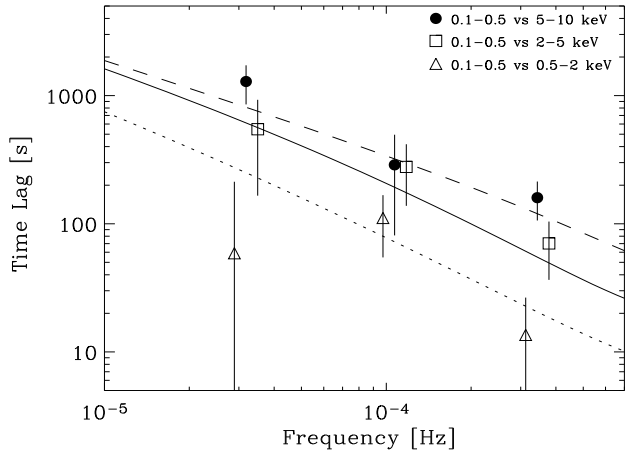,width=8.5cm,height=6.23cm,bbllx=72pt}
\caption {Time lags between the 0.1-0.5 keV band and 0.5-2, 2-5 and 5-10 keV bands 
(triangles, open squares and filled circles respectively), for NGC4051 \xmm\
data. The lines represent the model lags calculated using Eq. \ref{eqlags} with
model parameters fitted to match the observed
PSDs as described in the text. The
dotted, solid and dashed lines were calculated between synthetic light
curves with emissivity indices $\gamma_{\rm soft} = 2.5$ and
$\gamma_{\rm hard} = 3.3, 5.5$ and
$\infty$, and $\beta=1$, assuming a black hole mass of
$5 \times 10^{5}$~M$_\odot$. The filled circles are placed at the correct
frequency used for the three data sets, while
the squares and triangles have been shifted in this plot to slightly lower
and higher frequencies respectively, to avoid overlapping the error bars.}
\label{lags_NGC}
\end{figure}

The time lags in \ngc\ reported by \citet{McHardy4051} are among the
largest measured in AGN, with values of $\sim$ 0.25, 2.5 and 4.5
per~cent of the variability time-scale, when calculated between the
0.1-0.5 keV band and the 0.5-2, 2-5 and 5-10 keV bands,
respectively. The slopes of the lag spectra are not well constrained
but are consistent with -1. To reproduce these large lags with the
model, we chose a fairly extended soft emitting region, with
emissivity index $\gamma_{\rm soft}=2.5$, assumed
 $(H/R)^2\alpha =0.3$ at $r_{\rm min}=6$, used a value of $\beta=1$ to
produce a lag-spectrum slope $\sim-1$ and used $D=0$. Using
Eq. \ref{eq1}, we calculated the lag spectra for these model
parameters, choosing emissivity indices for the remaining energy bands
to match the observed lags. We used $Q=1$ for the calculation,
although other $Q$ values produce equally good fits as the width of
the input signal PSD has a very small effect in the lag spectrum over
the frequency range used here (see Sec. \ref{input}).  The resulting
lag spectra for $\gamma_{\rm hard}$=3.3, 5.5 and $\infty$ are shown in
Fig. \ref{lags_NGC}, in dotted, solid and dashed lines,
respectively. To convert light-crossing times to seconds, we assumed a
mass of $5\times 10^5 M_\odot$ \citep{Shemmer}.  Overlaid in this
figure, we show the lags measured from \xmm\ data for \ngc\ , where
the observed light curves were constructed as described in
\citet{McHardy4051}.  Although the errors on the data are large, it is
clear that the chosen model parameters are consistent with the
amplitude, slope and energy dependence of the measured lags.

\subsection{Energy dependence of the PSD}

\begin{figure}
\psfig{figure=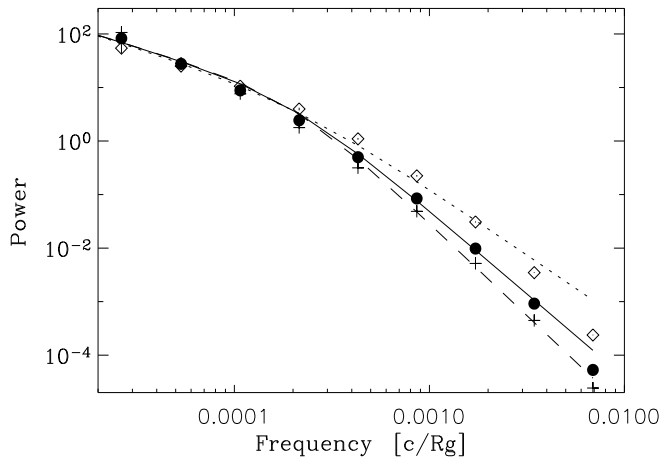,width=8.5cm,height=6.23cm,bbllx=60pt}
\caption {PSDs of three synthetic light curves of emissivity indices
$\gamma=2.5, 3.3 $ and 5.5 in crosses, filled circles and open
diamonds, respectively, calculated with $(H/R)^2 \alpha=0.3$, $D=0$, $r_{\rm min}=6$ and $Q=1$. The lines mark the bending power law model fit
to each PSD in the frequency range shown, with fixed $\alpha_{\rm
L}=-1.1$ and $f_{\rm b}= 2\times10^{-4} c/R_{\rm g}$. The PSD and
model normalisations have been rescaled to match at the low frequency
end. The PSD of model light curves can be fitted approximately with a
bending power law of fixed $\alpha_{\rm L}$ and $f_{\rm b}$, which
highlights the differences in the high frequency slope.}
\label{psd_surr_ngc} 
\end{figure}

\ngc\ was monitored with {\sl RXTE} and \xmm\ on long and short
time-scales respectively by \citet{McHardy4051}, producing a well
sampled PSD over 6 decades in frequency. The high frequency PSD in
\ngc\ shows strong energy dependence, varying from a slope $\alpha_{\rm
H}=2.91 \pm 0.12$ in the 0.1-0.5 keV band to $\alpha_{\rm H} = 2.06\pm
0.16$ in the 2-5 keV band.

The difference in emissivity indices for these energy bands, suggested
by the lag spectra, can also explain the energy dependence of the high
frequency PSD slope. To measure the energy dependence we produced
model light curves with broad input signal PSDs, $Q=0.5$, and model
parameters as described above and fitted the final PSD with a bending
power law model. To make a fair comparison with the data, we followed
the procedure used by \citet{McHardy4051} by fixing the 
low-frequency slope $\alpha_{\rm L}$
to -1.1 and the bend frequency $f_{\rm b}$ to the best fitting value for an
intermediate emissivity index, and fitted the simulated PSDs over a
similar range in frequencies.  Using again $\gamma_{\rm soft}=2.5$ for
the 0.1-0.5 keV and $\gamma_{\rm hard}=5.5$ for the 2-5 keV band, we
obtained PSDs that differ by $\Delta\alpha_{\rm H}=-0.95$, spanning a
similar range to that measured for the real data, $\Delta \alpha_{\rm
H, NGC}=-0.85\pm0.2$.  The resulting PSDs and bending-power law fits
are shown in Fig. \ref{psd_surr_ngc}, for emissivity indices $\gamma=$
2.5, 3.3 and 5.5. Notice that in this figure the PSD normalisations
have been rescaled to match at the low frequency end, to highlight the
different changes in slope. The filled circles in this figure
correspond to $\gamma = 3.3$, which would represent the 0.5-2 keV
band, as seen in Section~4.1. Correspondingly, the difference in high
frequency PSD slope between the $\gamma=2.5$ and $\gamma=3.3$ light
curves is $\Delta\alpha_{\rm H}=-0.34$, placing the $\gamma=3.3$ light
curve between the 0.5-1.0 keV ($\Delta \alpha_{\rm H, NGC}=-0.25$) and
1.0-2.0 keV ($\Delta \alpha_{\rm H, NGC}=-0.54$) bands.

To check the consistency of the time-scales, we compared the bend
frequency $f_{b, NGC}$ measured by \citet{McHardy4051}, using the
combined {\sl RXTE} and \xmm\ light curves, to model light curves
covering a similar range in frequencies. As the \ngc\ data used for
this fit corresponded to the 4--10 keV \xmm\ band, we used the `hard'
simulated light curves with $\gamma$ between 5.5 and $\infty$. The
best fit bend frequencies for $\gamma=5.5 $ and $ \infty$,
respectively, are $4.6\times 10^{-4}$ and $2 \times 10^{-3} c/R_{\rm
g}$. Equating these values to $f_{b, NGC}=5\times 10^{-4}$ Hz yields
a mass of $2 - 8\times10^5 M_\odot$, consistent with the \ngc\
reverberation mapping mass of $5^{+6}_{-3}\times10^5 M_\odot$
\citep{Shemmer}.

A similar energy dependence of the PSD has been found in other AGN. As
an example, the PSD of the Seyfert 1 galaxy \mcg\ can also be fit with
a bending power law model obtaining flatter high-frequency slopes for
higher energy light curves \citep{VaughanMCG}. The energy dependence
in this case is less pronounced than in \ngc, possibly suggesting that
the energy bands used have closer emissivity indices. Incidentally,
emissivity indices that are closer in value would also produce smaller
time lags, as observed in \mcg. However, more parameters contribute to
define the observed lags, for example, the mass of the central object
changes the frequency range probed in terms of the light crossing
times. A comparison of the fractional lags at, for example, the
PSD bend frequency in each case gives a mass-independent measure, so
an extrapolation of the lag spectra in \ngc\ published by \citet{McHardy4051}
down to the break time-scale would suggest that the lags in this case
are indeed systematically larger than in \mcg.
\section{Application to Cyg X-1 data}
\label{xrb}

We now apply our simple model to data from the well-studied,
persistent black hole XRB Cyg~X-1, in the high/soft state.

Our model produces light curves with a $\sim1/f$ PSD shape down to low
frequencies.  This is because we assume that the input signals
originate at many different radii throughout the accretion disc, are
equally separated in the logarithm of frequency and have identical rms
amplitudes of variability. PSDs of this shape are seen for several AGN
(see previous section) and Cyg~X-1 in its high/soft state. 

We will use data from a short (2.3~ks) {\it RXTE} observation of
Cyg~X-1 in the high/soft state, Observation ID 10512-01-09-01,
obtained on 1996~June~18\footnote{We do not combine data from
different high/soft state observations, since there are significant
variations in the timing properties in the high/soft state on
time-scales of hours or longer (\citealt{Axelsson,Cui97}). We choose
this particular observation as it is free of Lorentzian components
\citep{Cui97} and is also fitted by \citet{McHardy4051}.}.  Using
Proportional Counter Array (PCA) light curves with $2^{-8}$~s
resolution, we measured the lag spectrum and PSD in two bands,
2--5.1~keV (soft band) and 8--13~keV (hard band). We subtracted
Poisson noise, including the appropriate deadtime correction,
e.g. \citet{RevHifreq}, before calculating the ratio of the PSDs.  To
keep the fitting simple, we consider only the ratio of the PSDs, 
and not the energy-dependent PSD shapes themselves, so we do not
attempt to match the PSD shape by assuming any particular input
signal.  This approach allows us to estimate the lags and filtering
effect in a simple analytical way, by assuming that each frequency
contributing to the lag spectrum and PSD ratio corresponds to a single
radius, so input signal PSDs do not overlap in frequency, with the
lags and PSD filtering determined using the analytical expressions
given in Appendix~A. We note that \citet{McHardy4051} do not claim
any evidence of energy dependence of the PSD in their fitting of the
same data used here, but we do find a significant energy dependence
using the PSD ratio (see below).  The discrepancy may result from the
fact that the measured PSD ratio is, in fact, more sensitive than
independent fits to the PSD, because the light curves in different
bands are correlated, so that statistical scatter in the PSD due to
the stochastic nature of the light curves is in the same direction in
both bands, and its effect on the PSD ratio is therefore mitigated.

We first stress that due to the complexity of fitting even our 
relatively simple model
to the data, and the difficulty of quantifying certain measurement
uncertainties (see below), we will only test the broad consistency of
our model with the data.  Therefore, we will not quote statistical
errors on parameters derived here, which should be treated as only
indicative of the underlying physical parameters.  Since our model
contains many parameters with complex degeneracies between them, we
only fit the model with a few parameters left free, to demonstrate
whether the model can reproduce the data.  For all our fits, we
assumed a black hole mass of 10~M$_\odot$. The mass provides the
scaling factor between the units of light-crossing time-scale used in
our method, and the observed units of seconds.  The effect of changing
mass is therefore equivalent to that of a simple linear change in the
quantity $(H/R)^{2}\alpha$ (see Section~3.2) at all radii, and is
rather weak except at high frequencies. Consistent with our earlier
simulations, we initially keep $(H/R)^{2}\alpha=0.3$ at all radii. We
assume a fixed outer disc radius $r_{\rm max}=10^{5}$~$R_{\rm g}$, but
allow the inner disc radius $r_{\rm min}$ to vary, down to a minimum
value of 1.23~$R_{\rm g}$ (i.e. the minimum stable orbit for a Kerr
black hole). For simplicity, the soft emissivity index was initially
fixed at $\gamma_{\rm soft}=3$, while $\gamma_{\rm hard}$ was allowed
to vary.  We also introduced an additional normalising factor for the
lags, so that all the lags predicted by the model can be increased by
a single factor, which we leave free to vary in the fits. This
renormalisation represents the possibility that the fluctuations
propagate at some speed proportional to the theoretically predicted
local drift velocity (see section~2.1), but not necessary equal to it.

Note that the PSDs in different energy bands can have different
overall normalisations for a variety of reasons.  For example, if a strong
constant thermal emission component is present, as in the high/soft
state (see \citealt{Churazov}), the variability is diluted in the
soft band and the PSD normalisation reduced.  Alternatively, spectral
pivoting at high energies (e.g. due to Compton cooling of the
corona), which we do not model here, can lead to a larger
normalisation in the soft band compared to the hard band (e.g. see
\citealt{Uttley3227} for evidence of this effect in the AGN
NGC~3227).  For these reasons, when fitting our model we allow the
normalisation of the PSD ratio to be free.

\begin{figure}
\psfig{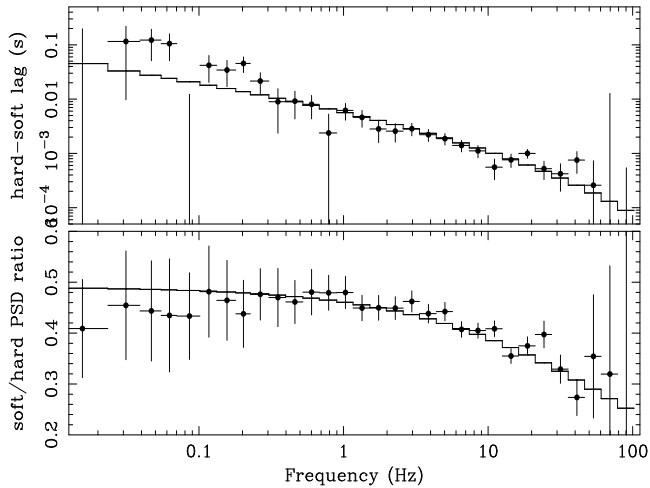}
\caption {Time lags (top) and soft/hard PSD ratio (bottom) as a
function of Fourier frequency for Cyg~X-1 in the high/soft state
(1996 June 18). The solid lines show the best-fitting analytical
model described in the text.} 
\label{cyghilagratio}
\end{figure}

The observed lag and soft/hard PSD ratio spectra are shown in
Fig.~\ref{cyghilagratio}, together with the best-fitting model
resulting from a joint fit of the lag and filtering equations to the
data, i.e. we assume the same parameters for fitting both the lag and
ratio plot. The reduced chi-squared, $\chi^{2}/{\rm d.o.f.}=0.81$,
indicates an excellent fit, for a renormalisation factor of 1.86
(i.e. fluctuations propagate at $\sim1.9$ times lower drift velocity
than given by standard disc theory), $r_{\rm min}=1.23$, and
$\gamma_{\rm hard}=3.53$.  We caution however that the goodness-of-fit
is almost certainly over-estimated, since the random errors in the
soft and hard PSDs are correlated at low frequencies due to the
intrinsic correlation between the light curves in these bands (the
correlation is much weaker at high frequencies due to Poisson noise
effects), and hence the error bars at these frequencies are almost
certainly over-estimated.  Furthermore, there is likely to be some
additional uncertainty in PSD ratio at high frequencies ($>20$~Hz), since an
assumed Poisson noise level had to be subtracted from each PSD before
the ratio was taken, and the uncertainty in the assumed noise level is
not accounted for in the fit. 

The small inner disc radius preferred by the data is consistent with a
rapidly rotating black hole.  However, this interpretation is strongly
governed by our assumptions about the emissivity profile, and our
simplifying assumption that each frequency corresponds to a single
input signal. Since the travel time of material from 6~$R_{\rm g}$ to
1.23~$R_{\rm g}$ is very short, the value of $r_{\rm min}$ is not
driven by lags at lower frequencies, but is mainly due to a
combination of the relatively large high-frequency lags, and the weak
drop in PSD ratio at high frequencies.  Indeed, the frequencies
corresponding to signals at 6~$R_{\rm g}$ and 1.23~$R_{\rm g}$ are
65~Hz and 700~Hz respectively, so that if the edge of the emitting
region were observed at 6~$R_{\rm g}$ a very pronounced drop in both
the lags and the PSD ratio would be observed approaching this
frequency, hence the smaller radius is preferred.  However, if the
input signals overlap substantially, the larger lags from signals at
larger radii can contribute even at high frequencies, so a strong drop
in lag and PSD ratio would not be expected (see section
\ref{input}). There may also be substantial emission at radii smaller
than $r_{\rm min}$, which is not considered in our model.  This would
also increase the lags at high frequencies and reduce the effects of
filtering which cause the drop in PSD ratio at high frequencies.
Thus, the small value of $r_{\rm min}$ inferred from our fit is quite
model-dependent and should be treated with caution.

The fact that the model can explain the soft/hard
PSD ratio also implies that it can explain the results of
Fourier frequency-resolved spectral analysis of the high/soft state
continuum, as presented by \citet{Gilfanov2000}.  The Fourier-resolved
spectra become harder at higher temporal frequencies, with the change
in photon index consistent with the frequency-dependent change in PSD 
ratio which we measure here.

Finally, for completeness we note here that, unlike the smooth $1/f$ shape
seen in the high/soft state,
Cyg~X-1 in the low/hard state is characterised by a band-limited
PSD with a few humps, which can be modelled as four broad Lorentzian
components \citep{Pottschmidt03}. The lag spectrum also reveals a
stepped structure, but it still follows approximately a $f^{-0.7}$
power law slope. Although the model can fit reasonably well the
general trend seen in the PSD ratio and lag spectra, it cannot
reproduce these detailed features, indicating that some of the model
assumptions are not appropriate in this case. In particular, our
simple implementation of the model uses power law emissivity indices
and equal variability power per contributing annulus, so these smooth
functions cannot produce additional structure at a few given
frequencies. The general scheme of propagating fluctuations can still
be correct but additional assumptions need to be made, for example,
assuming that a few annuli have much enhanced variability power will
certainly produce a bumpy PSD and might reproduce the steps in the lag
spectra, as suggested by \citet{Nowak00} and also \citet{Kotov}.
\section{Discussion}
\label{discussion}

The extended emitting region in our model is responsible for the high
frequency bend in the PSD, and the associated radial emissivity
profiles produce the energy dependence of the PSD and the time
lags. In this section we will summarise the spectral-timing properties
produced by the model, consider the implications of the model for the
size of the emitting region, and discuss some possible improvements.

\subsection{Time lags}

Keeping the local variability time-scales tied to the propagation
time-scale produces time lag spectra of power law slope $\sim -1$ or
flatter, and lags of $\sim 1 -10\%$ of the variability time-scale, for
a wide range of model parameters. As discussed in Sections \ref{agn}
and \ref{xrb}, these simple assumptions produce lag spectra that match
the data well in time-scale dependence and amplitude. We note once
again that these lags arise solely due to the difference in emissivity
profiles of the X-ray energy bands and do not involve any other spectral
evolution of the emitting region. The amplitude of the lags depends
mostly on the emissivity indices of the energy bands, increasing
rapidly with their difference, up to $\Delta \gamma \sim 1$, above which
the lag values tend to saturate. Significant lags can appear
between energy bands characterised by similar emissivity profiles and,
correspondingly, similar PSD shapes. In particular, the PSD of Cyg~X-1
in the high/soft state shows weak energy dependence (see 
Section~5.1), but the PSD ratio
and lags can still be reproduced simultaneously with close emissivity
indices for each energy band ($\Delta \gamma \sim 0.5$), 
provided that the propagation time-scale
is slightly longer than the local fluctuation time-scale, and assuming
that the inner disc radius is small. The power-law
shape of the lag spectra is quite robust, its slope depends only
weakly on disc structure parameters $(H/R)^2\alpha$. Incidentally, the
stability of the lag spectra might explain the behaviour seen in,
e.g. Cyg~X-1, in different spectral states. This object shows very
different PSD and energy spectra in the high/soft and low/hard states,
indicative of different disc configurations, but surprisingly similar
lags \citep{Pottschmidt}.

\subsection{PSD shape}

A comparison with AGN and BHXRB data in the high/soft state shows that the
filtering effect of the extended emitting region, acting on a simple
$1/f$ underlying PSD shape, can broadly reproduce their PSD shape and
energy dependence. Cyg~X-1 in the low/hard state, however, requires a
more complex \emph{intrinsic} PSD. In either case, the extended
emitting region introduces a bend in the PSD in addition to any
intrinsic curvature, and produces the energy dependence of the
filtered PSDs.

Note that the bending power-law, used to fit AGN data, is only an
approximation to the actual PSDs produced by the fluctuating-accretion
model. The filtered PSD bends down continuously at high frequencies
and has no well-defined high-frequency power law slope. The difference
between the bending power-law model and the filtered PSDs can be
appreciated in Fig. \ref{psd_surr_ngc}, where the single-bend PSD fit
overestimates the power in the highest frequency bins. However, as
this region of the PSDs from real data is often heavily affected by
Poisson noise, the associated error bars and scatter are large and it
is not possible to appreciate any deviations from a simple power law
slope. Therefore, even better AGN data would be needed to discern if
the model can replicate the exact PSD shape or if additional
variability components are needed.  We also note here that, although
\citet{RevHifreq} fit the high/soft state PSD from combined data from
1996 June 4-18 with a simple power law (index -2.1) up to
$\sim200$~Hz, the signal-to-noise at these frequencies is still fairly
low, and the PSD during that time is known vary in shape between
observations \citep{Cui97}, which makes interpretation of the
underlying shape even more difficult.  \citet{Axelsson} have recently
demonstrated that the high/soft state PSD can be fitted with weak
Lorentzians {\it in addition} to a $\sim 1/f$ power-law component with
an exponential cut-off. This last component is reminiscent of our
simulated PSDs. Therefore, our model may be considered as 
representative of the times when the Lorentzians in the PSD are very weak or
absent.

In our model, thick disc parameters, $\alpha (H/R)^2=0.3$, in the
inner regions of the accretion flow and inner radius $r_{\rm min}= 6$
put the break frequency around $10^{-4}-10^{-3}c/R_{\rm g}$
(i.e. $\sim$2 - 20 Hz for a $10 M_{\odot}$ black hole, $2\times
10^{-5}-2\times 10^{-4}$ for a $10^6 M_{\odot}$). These values are in
general agreement with the average break frequencies of AGN and
BHXRBs, indicating that fluctuations on the viscous time-scales of a
geometrically thick accretion flow are appropriate to explain the
variability. A geometrically thick disc is necessary to produce the
observed high frequency fluctuations as \emph {viscous}
frequencies. If the fluctuations were instead produced on dynamical
time-scales, or a magnetic time-scale related to this, as in
e.g. \citet{King}, then a thinner flow might be allowed. However, the
fluctuations need not only be produced but also propagated, which
poses a difficulty for much thinner accretion flows, as fluctuations
on time-scales much shorter than the propagation time-scale are easily
damped \citep{Churazov}. This, of course, does not rule out an
additional thin disc, possibly underlying the thick flow, that might
contribute to the flux but not to the variability.

\subsection{Extent of the emitting region}

In our implementation of the variability model, the X-ray emitting
region extends out to large radii. The bend in the PSD is produced by
the radial distribution of variability time-scales and emissivity
profiles alone and is not related to a characteristic time-scale at
the maximum radius of emission. Therefore, an outer edge to the
emitting region might be allowed but is not required. The steep
emissivity profiles concentrate most of the emitted flux in the centre
so it would make little difference to the PSD if the emitting region
were truncated at some radius or if it extended out to
infinity. However, the extent of the emitting region has a major
impact on the lag spectra. If, for example, the X-ray emission is
confined to the central few $R_g$ and responds to the variability
pattern produced outside this radius, then time lags would still
appear due to the radial segregation of the energy bands but, to first
order, the lags would have the same value at \emph{all} Fourier
frequencies \citep[e.g.][]{Nowak_lags}.  In our case,
frequency-dependent lags appear because the fluctuations are produced
\emph{within} the emitting region. This consideration gives further
support to the idea that $f_{\rm b}$ does not correspond to the
characteristic frequency of an outer edge of the emitting region, as
noted above, as these lags are observed far below the break
frequency. This, in turn, implies that the X-ray emitting region
should extend at least as far as the radius corresponding to the
longest time-scales where frequency-dependent lags are observed.

\subsection{Improvements to the model}

We have made a number of simplifying assumptions that seem to
reproduce the broad spectral timing properties
of AGN and BHXRBs. The detailed
predictions of PSDs and lag spectra can be improved by including some
additional effects.

As a first step, the assumption that all radii in the disc produce the
same amplitude of fluctuations can be relaxed, allowing for strongly
enhanced variability at a few discrete annuli. This scenario might
represent Cyg~X-1 in the low/hard state where the PSD appears to
be composed of a few broad Lorentzians, so it would be interesting to
check if the effect of the extended emitting region would reproduce
well the steps observed in the PSD ratio and in the lag spectra.

To simplify the calculations, we have only allowed for power-law
radial emissivity profiles. Though this appears to be a reasonable
assumption given the profile of total energy loss expected for an
accretion disc, it is not clear that each energy band should follow a
simple power-law profile.  For example, the energy spectrum reveals
other components in addition to the direct continuum that might be
expected from an optically thin X-ray emitting corona.  In BHXRBs in
the high/soft state, black body disc emission is also seen, although it
does not vary significantly on short time-scales \citep{Churazov}, so
may not contribute to the observed PSD and lag spectra.  However, in
several AGN (e.g. NGC~4051, \citealt{Uttley4051}), strongly variable
soft excess emission is seen below $\sim1$~keV, which may be
associated with Comptonised disc emission and hence could contribute a
different component to the emissivity profile in the soft band.  Other
complex profiles might be expected. For example, by fitting the
profile of the broad iron line in \mcg\ with a relativistic disc-line
model, \citet{Fabian} have inferred a broken power-law emissivity
profile, which steepens from $\gamma=2.5$ (i.e. similar to the value
we infer for NGC~4051) to $\gamma=4.8$ within 6~$R_g$ of the black
hole. Thus, more complex emissivity profiles, perhaps with breaks or
additional components, and taking account of general relativistic
effects, could be considered.  Additional variable components, arising
for example through reflection or reprocessing of the primary variable
emission, can also affect the coherence and time lags and should be
included in more detailed variability models. In particular, in high
quality data from Cyg~X-1, \citet{Kotov} show that a reflection
component is necessary to reproduce the detailed \emph{energy
dependence} of the lags.

The simple fluctuating-accretion model presented here agrees generally
well with the observational variability measurements, but still needs
to be set on physical grounds to be fully validated. A physical
treatment must consider two things, the source of the variability
(i.e. source of accretion rate fluctuations, for example), and the
source of emission. Work on the evolution of a physical (if
simplified) accretion flow, implemented with numerical simulations
\citep[e.g][]{King}, indicate that the necessary accretion rate
fluctuations and wave-like propagation on viscous time-scale can be
realised physically. Finally, a model for the emission process, and
its coupling to the accretion rate fluctuations, will be necessary to
provide a physical motivation for the radial emissivity profiles of
each energy band and, with it, a meaningful interpretation of the
emissivity indices.

\section{conclusions}
\label{conclusions}
We have used a computational fluctuating-accretion model to reproduce the
spectral-timing properties of X-ray light curves from AGN and
BHXRBs. In the model, the variability is produced throughout the
accretion flow as perturbations of the accretion rate on the local
viscous time-scale, produced by many signals which are geometrically
spaced in temporal frequency.  The fluctuations propagate inward at
the radial drift velocity and modulate the fluctuations produced
further in. This general scheme, introduced by \citet{Lyub}, produces
light curves which possess $1/f$ type PSDs over a broad range of
time-scales and follow a linear rms-flux relation. Following
\citet{Kotov}, X-ray emission from a radially extended region is
assumed to track the local accretion rate fluctuations and to show a
power law radial emissivity profile, which steepens for higher energy
bands. The different emissivity profiles and the inward propagation
produce differences between the PSDs of each energy band and introduce
time lags. Our implementation of the model is 
successful in reproducing the observed variability properties of AGN
and BHXRBs, as follows:

\begin{enumerate}
\item We reproduce the expected $\sim1/f$ 
frequency-dependence of time lags between energy bands, 
and lag amplitudes of a few per cent of the variability time-scales,
predicted by \citep{Kotov} and observed in BHXRB and AGN 
variability data.

\item We demonstrate that the amplitudes and slopes
of the lag spectra are only weakly dependent on
changes in the emissivity indices, the radial dependence
of the product of disc parameters $(H/R)^2 \alpha$, and
the strength of damping of accretion fluctuations. 
The robust nature of the lag spectra
may help to explain the similarity of lag spectra in the low/hard and
high/soft states in Cyg~X-1 \citep{Pottschmidt},
despite the very different PSD
and energy-spectral shapes in these states.

\item The extended emitting region suppresses short time-scale
fluctuations, producing a gradual bend in the PSD at high
frequencies. The `filtered' PSD shapes resemble those of AGN and
BHXRBs in the high/soft state (e.g. \citealt{McHardy4051}) and can be
approximately fitted with a bending power-law model.  The bend
frequencies obtained match the observed frequencies if the
fluctuations are produced on the viscous time-scales of a
geometrically thick accretion flow. Values of $(H/R)^2\alpha \sim 0.3$
gave adequate bend frequencies compared to data from Seyfert galaxy
\ngc\ and the BHXRB Cyg~X-1.

\item The fact that all energy bands are produced by the same emitting
region can maintain a high coherence ($>0.95$) between light curves of
different energy bands, as observed by e.g. \citet{Nowak_lags},
\citet{VaughanMCG}. However, the coherence can be substantially
reduced at high frequencies if input signals at each radius have
fairly broad PSDs and the difference between emissivity profiles in
different bands is large. Coherence is also reduced if damping of the
inward-propagating fluctuations is significant.  These results may
provide an explanation for the drops in coherence observed at high
frequencies in various BHXRBs and AGN \citep{Nowak_lags,
VaughanMCG,McHardy4051}.

\item The model can simultaneously reproduce both the observed
energy-dependence of PSD shape and the lag spectrum for the AGN
NGC~4051 and the BHXRB Cyg~X-1 in the high/soft state, although the
inferred propagation times in Cyg~X-1 are slightly slower than the
drift velocity expected for a standard disc.  In the low/hard state of
Cyg~X-1, our simple model can reproduce the general slope of the lag
spectrum but cannot account for the stepped shape of the lag
spectrum. Modelling the detailed features of this spectrum would
require additional assumptions about the mechanism producing the
variability.

\item Direct comparison with spectral-timing data from AGN and BHXRBs
shows good general agreement with the model. However, we note that if
the model is correct, the empirical bending or broken power-law models
used to fit the PSDs of AGN and BHXRBs in the high/soft state must
break down for high signal-to-noise data, because they only
approximate the gradual bending shape predicted by our model. Also, we
note that the model implies that the PSD bend time-scale measured for
AGN and BHXRBs is a function of both the underlying PSD of accretion
fluctuations (which can contain intrinsic breaks) {\it and} the
emissivity profile of the observed energy band (which introduces a
high-frequency bend).  Therefore, the PSD bend time-scale is not
necessarily associated with the outer radius of the emitting region,
and this region can, in principle, be extremely large, provided the
emissivity profile is relatively steep.

\end{enumerate}

\section*{Acknowledgments} 
This research has made use of data obtained from the High Energy
Astrophysics Science Archive Research Center (HEASARC), provided by
NASA's Goddard Space Flight Center.  We wish to thank the two
anonymous expert referees for their valuable comments.  PA
acknowledges support from the International Max-Planck Research School
in Astrophysics (IMPRS), travel support from grants associated with
the NASA {\it RXTE} Guest Observer program, and the hospitality of
NASA's Goddard Space Flight Center.  PU acknowledges a Research
Associateship from the National Research Council.

\label{lastpage}
\appendix

\section[]{ Analytical estimates for filtered PSD and lags}

As mentioned in Sec. \ref{model}, the extended emitting region acts as a
low-pass filter on the PSD. This filter factor, $F$, can be approximated as:

\begin{equation} {\rm PSD}_{\rm filt}(f)={\rm PSD}(f)\times F(r_f)
={\rm PSD}(f) \left(\frac{\int^{r_f}_{r_{\rm min}}\epsilon (r)2 \pi r dr}
{\int^{\infty}_{r_{\rm min}}\epsilon(r)2 \pi r dr}\right)^2
\end{equation} 

Taking the emissivity profile $\epsilon(r)=r^{-\gamma}(1-\sqrt{r_{\rm min}
/r})$, the filter factor takes the form (for $\gamma>2$):

\begin{eqnarray}
F(r_f)=\left( \frac{\frac{r_f^{2-\gamma}-r_{\rm min}^{2-\gamma}}
{2-\gamma}-\sqrt{r_{\rm min}}\frac{r_f^{1.5-\gamma}-r_{\rm min}^
{1.5-\gamma}}{1.5-\gamma}}{ \frac{r_{\rm min}^{2-\gamma}}
{\gamma-2}+\frac{r_{\rm min}^{2-\gamma}}{1.5-\gamma} }\right) ^2
\nonumber\\
=\left( 1-\frac{2 r_f^{2-\gamma}}{r_{\rm min}^{2-\gamma}}
\left( \gamma-1.5 - \sqrt{\frac{r_{\rm min}}{r_f}}(\gamma-2) \right)
\right) ^ 2 
\end{eqnarray}
In our implementation, the fluctuations are produced on the viscous time-scale
at the radius of origin, so $r_f= (2\pi f(R/H)^2/\alpha )^{-2/3}$.
Substituting this value in the equation above gives the PSD filtering factor as a
function of frequency $f$, where $f$ is in units of $c/R_g$. This approximation
for the filtering effect is appropriate for narrow-input-PSD signals, where most
of the variability power produced by each annulus is produced on the
corresponding viscous time-scale. Finite travel time effects contribute to cancel
out variability produced at higher frequencies than $f_{\rm visc}$, so the
filtered PSD in this case is distorted further. 

For the calculation of the time lags, the average travel time of a signal
produced at $r_s$ can be approximated as:

\begin{equation}
\bar{\tau}(s)=\frac{\int^{r_s}_{r_{\rm min}}\tau(r,r_s) \epsilon (r) 2\pi r dr}
{\int^{r_s}_{r_{\rm min}} \epsilon(r) 2 \pi r dr}, \tau (r,r_s)=
\int_{r}^{r_s}{\frac{d\tilde r}{v(\tilde r)}}
\end{equation}

If the fluctuations propagate with viscous velocity, $v(r)=\alpha(H/R)^2 
r^{-1/2}$, then $\tau (r,r_s)=\frac{2}{3 \alpha(H/R)^2}(r_s^{3/2}-
r^{3/2})$ and the average travel time can be calculated using the emissivity 
profile given above. For example, for the case where $\gamma \neq 1.5, 2, 
3, 3.5$, the average travel time is:  

\begin{eqnarray}
\bar{\tau }(s)= \left[ \frac{2}{3\alpha (H/R)^2} 
\right( r_s^{1.5}\frac{r_s^{2-\gamma }-r_{\rm min}^{2-\gamma }} 
{2-\gamma}- r_s^{1.5} r_{\rm min}^{0.5} 
\frac{r_s^{1.5-\gamma }-r_{\rm min}^{1.5-\gamma }} {1.5-\gamma } 
\nonumber \\ 
- \frac{r_s^{3.5-\gamma }-r_{\rm min}^{3.5-\gamma }}{3.5-\gamma } +
 r_{\rm min}^{0.5} \frac{r_s^{3-\gamma }-r_{\rm min}^{3-\gamma }}
{3-\gamma } \left) \right] \nonumber\\
/ \left[ \frac{r_s^{2-\gamma }-r_{\rm min}^{2-\gamma }}{2-\gamma }
-r_{\rm min}^{0.5}\frac{r_s^{1.5-\gamma }-r_{\rm min}^{1.5-\gamma }}
{1.5-\gamma } \right]
\end{eqnarray}
Substitution of the exponentials with logarithms in the appropriate terms 
give the equation for the rest of the cases. For signals produced on the viscous 
time-scale of the radius of origin, $r_s$ can be substituted by $r_f$ as above, 
producing average travel times as a function of temporal frequency. Finally, 
the difference between $\bar \tau(f)$ calculated for two different emissivity 
indices $\gamma$ gives the time lag spectrum corresponding to this 
$\gamma$ pair.

\section[]{Spectral-timing measurements}

We estimate the PSD using the periodogram, where the variability
power, $P(f_i)$, is calculated for discretely sampled frequencies $f_i$ as:
\begin{equation}
P(f_{i})=\frac{2}{\bar s ^2}\frac{\Delta t}{N}|{\rm Re}^2_S(f_{i})+
{\rm Im}^2_S(f_{i})|,
\end{equation}
where ${\rm Re}_S(f_{i})$ and ${\rm Im}_S(f_{i})$ are the real and
imaginary parts of the discrete Fourier transform of the time series
$s(t)$, $\bar s$ is the average count rate, $\Delta t $ is the sampling time
interval and $N$ is the number of sampled points \citep{Press}.

The Cross Spectrum between two time series, e.g. simultaneous soft and
hard light curves, $s(t)$ and $h(t)$, is defined as $C(f)= S^*(f)H(f)$,
where $S(f)$ and $H(f)$ are the Fourier transforms of the respective light
curves. The Cross Spectrum is a complex-valued function, from where the
coherence and phase lags can be extracted. The coherence $\gamma ^2$
for discretely sampled time series is calculated as:   

\begin{equation}
\gamma ^2 (f_{i})=\frac{ \langle {\rm Re}_C (f_{i}) \rangle ^2 +
\langle {\rm Im}_C (f_{i}) \rangle ^2}{\langle | {\rm S}(f_{i})|^2
\rangle \langle | {\rm H}(f_{i})|^2\rangle }
\end{equation} 
where ${\rm Re}_C (f_{i})$ and ${\rm Im}_C (f_{i})$ are the real and
imaginary parts of the Cross Spectrum $C(f)$ 
and angle brackets represent
averaging over independent measurements 
(i.e. the numerator
in the equation is the modulus-squared of the averaged cross-spectrum).
Note that here, the averaging of
independent measurements of the cross-spectrum and
PSD is carried out by binning up in frequency, as well as by averaging
measurements of the cross-spectrum measured at the same frequency
from separate light curve segments.

The argument of the Cross Spectrum defines the phase lags:
$\phi( f_{i})=\arg{\langle C(f_{i})\rangle}$, and from here the time lags,
$\tau(f_i)$, are calculated as:

\begin{equation}
\tau (f_{i})= \frac{\phi(f_{i})}{2\pi f_{i}}= \frac{1}{2\pi f_{i}} 
\arctan \left\{ \frac{ \langle {\rm Im}_C (f_{i}) \rangle }
{\langle {\rm Re}_C (f_{i})\rangle } \right\}
\end{equation} 

For a more detailed discussion on the measurement and interpretation
of coherence and time lags, and the determination of errors on these
measurements, see \citet{Vaughan_coh}, \citet{Nowak_lags} and, in the
context of AGN, \citet{VaughanMCG}.

\end{document}